\documentclass[aps,prb,reprint,superscriptaddress]{revtex4-1}

\pdfoutput=1
\usepackage{amssymb}
\usepackage{mathrsfs}
\usepackage{amsmath,amsfonts,bm,graphicx}
\usepackage{upgreek}
\usepackage{color}
\usepackage{etoolbox}

\newcommand{\app}[1]{Appendix~\ref{#1}}
\usepackage[colorlinks=true,linkcolor=blue,urlcolor=blue,citecolor=blue]{hyperref}

\newcommand{\fig}[1]{Fig.~\ref{#1}}
\newcommand{\Fig}[1]{Figure \ref{#1}}
\newcommand{\eq}[1]{Eq.~(\ref{#1})}

\newcommand{\Eq}[1]{Equation~(\ref{#1})}

\newcommand{\filetraja}{traja}
\newcommand{\filetrajb}{trajb}
\newcommand{\fileDv}{Dv}
\newcommand{\filealphaT}{alphaT}
\newcommand{\filepret}{pret}
\newcommand{\filemsd}{msd_v010}
\newcommand{\fileDT}{DT}
\newcommand{\fileFs}{Fs_10}
\newcommand{\fileFsKWW}{FsKWW_10}
\newcommand{\filebetaT}{betaT_10}
\newcommand{\fileDtalphaT}{DtalphaT_10}
\newcommand{\filechi}{chi4_10}
\newcommand{\filebondse}{bonds_e.pdf}
\newcommand{\filebondsp}{bonds_p.pdf}

\newcommand{\rv}{{\bf r}}
\newcommand{\phivconst}{0.01}
\providecommand{\mainsection}[1]{\section{#1}}

\newcommand{\phiv}{\phi_v}
\newcommand{\Vij}{V_{ijs_is_j}}

\newcommand{\Vkl}{V_{ijkl}}
\newcommand{\expV}{e^{-\upbeta\Vij} }

\newcommand{\av}[1]{\left\langle #1 \right\rangle}

\newcommand{\s}{{\{s_i\}}}
\newcommand{\sP}{{\{s_i>0\}}}
\newcommand{\n}{{\{n_i\}}}
\newcommand{\Zn}{Z_{\n}}
\newcommand{\ava}[1]{\av{#1}_{a}}
\newcommand{\avaZn}{\ava{{Z}_\n}}
\newcommand{\avq}[1]{\overline{#1}}
\newcommand{\Beta}{{\bm{\beta}}}
\newcommand{\sumsiN}{\sum_{~\{s_i>0\} \in \mathcal{P}_N~}}

\newcommand{\barM}{\bar{M}}
\newcommand{\fij}{\xi_{ij}}

\begin{document}
\title{Emergent facilitation behavior in a distinguishable-particle lattice model of glass}

\author{Ling-Han Zhang}
\altaffiliation[Present address: ]{Department of Physics, Carnegie Mellon University, Pittsburgh, Pennsylvania 15213}
\author{Chi-Hang Lam}
\email[Email: ]{C.H.Lam@polyu.edu.hk}
\affiliation{Department of Applied Physics, Hong Kong Polytechnic University, Hong Kong, China}
\date{\today}

\begin{abstract}
We propose an interacting lattice gas model of structural glass characterized by particle distinguishability and site-particle-dependent random nearest-neighboring particle interactions. This incorporates disorder quenched in the configuration space rather than in the physical space. The model exhibits non-trivial energetics while still admitting exact equilibrium states directly constructible at arbitrary temperature and density. The dynamics is defined by activated hopping following standard kinetic Monte Carlo approach without explicit facilitation rule. Kinetic simulations show emergent dynamic facilitation behaviors in the glassy phase in which motions of individual voids are significant only when accelerated by other voids nearby. This provides a microscopic justification for the dynamic facilitation picture of structural glass.
\end{abstract}

\maketitle

\mainsection{Introduction}
\label{sec:introduction}
Glassy dynamics still admits many open questions despite decades of intensive studies \cite{biroli2013review,stillinger2013review,berthier2011review}. When supercooled below the glass transition temperature $T_g$, many liquids can be quenched into the glassy phase, an amorphous solid-like state without long-range order.
Molecular dynamics (MD) simulations are able to capture the dramatic slowdown \cite{kob1995,kremer1990}, but a thorough understanding of the simulated dynamics also proves challenging. The study of simplified lattice models \cite{
edwards1975,kirkpatrick1987,fredrickson1984,palmer1984,kob1993,biroli2001,darst2010,lipson2013}
is thus important.
In particular, the p-spin model \cite{kirkpatrick1987} has inspired the random first-order transition (RFOT) theory \cite{kirkpatrick1989,kirkpatrick2014}, a leading theory of glass.
A potential issue in the p-spin model however is that it assumes externally imposed quenched disorder rather than the expected self-generated disorder,
although a density functional Hamiltonian with self-generated disorder has also been used to demonstrate RFOT  \cite{kirkpatrick1989SGD}.
Another promising theory is dynamic facilitation \cite{ritort2003review,garrahan2011review,chandler2010review} founded on kinetically constrained models (KCM) \cite{fredrickson1984,palmer1984}. An important example is a spin-facilitation model by Fredrickson and Andersen (FA) in which defects interpreted as low-density regions are allowed to evolve only when facilitated by the presence of adjacent defects \cite{fredrickson1984}.
A full microscopic justification of the facilitation rules still remains a challenge.

In this work, we formulate a distinguishable-particle lattice model (DPLM), which is a lattice gas model with effectively infinitely many particle-types. 
This generalizes other multi-species models for glass \cite{kob1995,darst2010,sasa2012}. It also models glassy systems in which most particles have distinct properties including polymers \cite{kremer1990},  polydispersive colloidal systems \cite{hunter2012} and monodispersive systems in which particle interactions admit random positional shifts \cite{mari2009,mari2011,charbonneau2014}. More generally, it is suggested to model also identical-particle glassy systems in which distinct particle properties effectively account for the positional disorder of particles at sub-lattice resolutions. 
DPLM  can be simulated at arbitrary temperature and particle density realizing physical systems ranging from dilute gases to glasses. Interestingly, the glassy phase exhibits dynamic facilitation as an emergent property.

\begin{figure}[tb]
\includegraphics[width=0.45\linewidth]{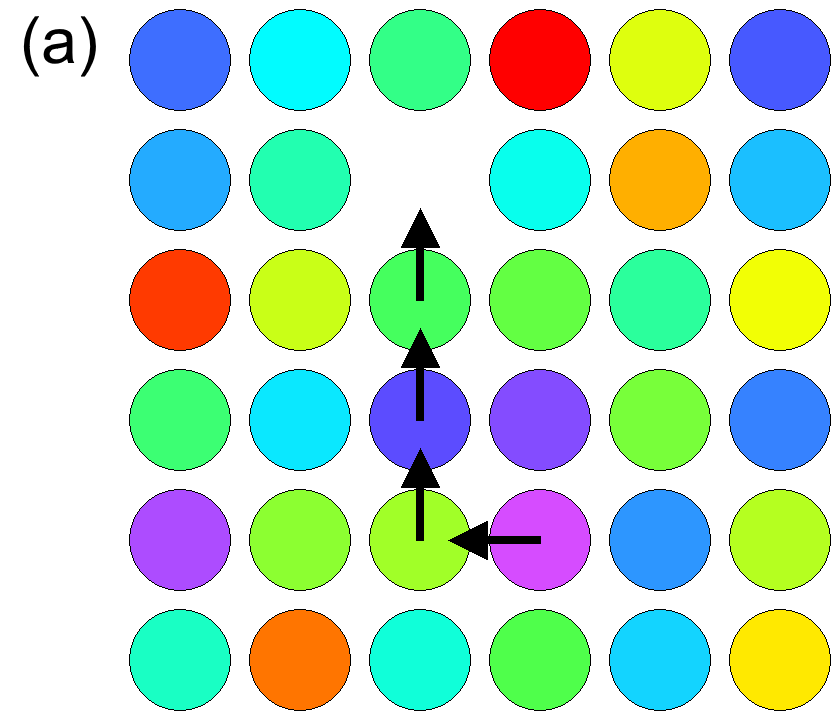}
~
\includegraphics[width=0.45\linewidth]{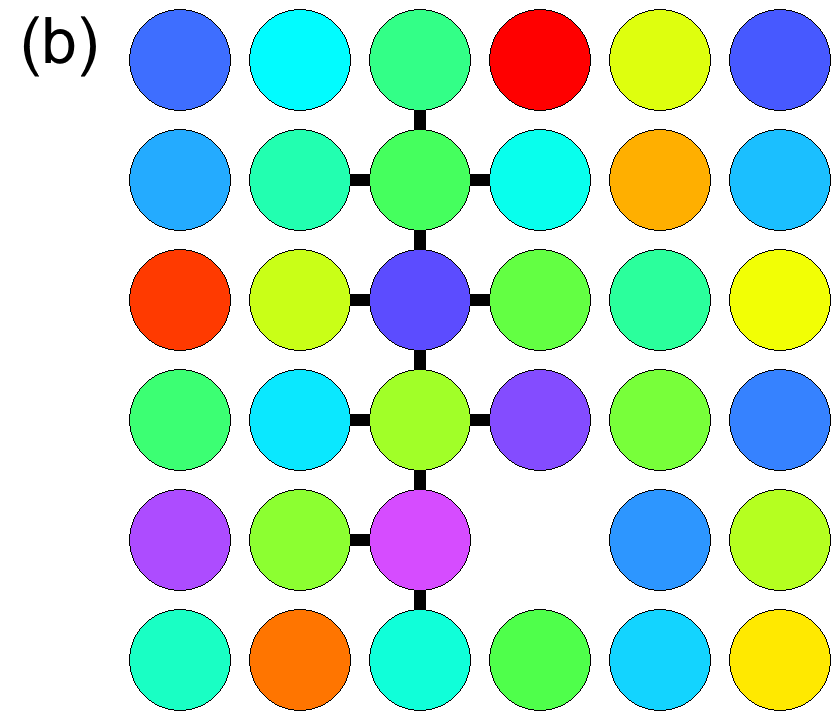}
\caption{
(a) Schematic diagram of a region with distinguishable particles randomly colored. The arrows indicate a possible sequence of hops by four particles arranged in a line.
The dynamics is equivalently described by four hops of a single void in the reversed direction. (b)
The particle displacements alter the nearest neighbor pairings and hence the pair interactions (indicated by black lines) along the $whole$ path. 
}
\label{fig:lattice}
\end{figure}

\mainsection{Model}
\label{sec:model}
DPLM is defined by $N$ particles on a 2D square lattice of unit lattice constant and size $L^2$ with periodic boundary conditions. No more than one particle can occupy each site.
Each particle is distinguishable from the others (see \fig{fig:lattice}).
For an occupied site $i$, the particle index $s_i = 1,2,\dots, N$ denotes which particle is at site $i$. For convenience, we let $s_i=0$ if the site is unoccupied, i.e. occupied by a void. The occupation number $n_i$ is hence 
\begin{equation}
  \label{n}
    n_i = 1 - \delta_{s_i,0}  
\end{equation}
where $\delta$ is the Kronecker delta. The whole set of $s_i$, rather than ${n_i}$, uniquely specifies the state of our system.

The total system energy is defined as
\begin{equation}
  \label{E}
   E = \sum_{<i,j>} \Vij n_i n_j
\end{equation}
where the sum is over all nearest neighboring (NN) sites. It can be equivalently written as 
\begin{equation}
  \label{E2}
   E  = \sum_{<i,j>'} \Vij
\end{equation}
where the sum is restricted to bonded NN sites $i$ and $j$, i.e. with both sites occupied by particles. 

A key feature is the site-particle-dependent interaction energy $\Vkl$. Its dependence on particle indices $k$ and $l$ means that each particle defines its own interaction strengths and this will be justified further. Effectively, each particle is a type of its own
generalizing multi-species models. 
In DPLM, each $\Vkl$ is time-independent and is an independent variable following a probability distribution $g(\Vkl)$ except when the symmetry $V_{ijkl}=V_{jilk}$ applies. We expect $\Vkl$ to be bounded below as in typical two particle interactions and thus $g(\Vkl)$ should not be for example a simple Gaussian.
For simplicity, $g(\Vkl)$ is assumed to be the uniform distribution in $[-0.5, 0.5]$ which leads to a particle interaction slightly attractive on average.

To better understand the time-dependence of the interactions, it is instructive to write \eq{E2} as 
\begin{equation}
  \label{E3}
   E  = \sum_{<i,j>'} V_{ij}(t)
\end{equation}
where $V_{ij}(t) \equiv \Vij$. 
We emphasize that while each interaction $\Vkl$ for any given sites $i$ and $j$ and particles $k$ and $l$ is a quenched random variable, the interaction $V_{ij}(t)$ at site $i$ and $j$ and arbitrary particles is $not$ quenched. Instead, $V_{ij}(t)$ admits an implicit time dependence via $s_i$ and $s_j$, which are time dependent and change in values when a particle at $i$ or $j$ is replaced. 
Equally importantly, $V_{ij}(t)$ has no explicit time dependence. A previous value can thus be exactly reinstated whenever a previous local particle configuration as specified by $s_i$ and $s_j$ is restored via the return of the particles. We believe that such particle-dependent local interactions with persistent memory capture essential characteristics of structural glass. A further subtle point is that since $V_{ij}(t)$ depends on time, the disorder in our model is $not$ quenched in the physical space, unlike spin-glass models \cite{edwards1975}. Instead, because of the time independence of $\Vkl$ and that the same interaction energy always applies to the same local particle configuration, the disorder is quenched in the configuration space. 

This site-particle dependence in $\Vkl$ is not necessarily due to possible diverse particle properties. Instead, it effectively account for the impacts of positional disorder at sub-lattice resolutions which are usually truncated in lattice models. A particle at site $i$ in a spatially disordered system in principle admits a small random offset $\Delta \rv_{i}$ from the exact lattice point.
This results in a random deviation in the atomic separation $\rv_{ij}$ between the particles at sites $i$ and $j$ and hence also in the pair interaction $\Vkl$.
Rather than explicitly modeling the disorder in $\Delta \rv_{i}$ or $\rv_{ij}$, we directly consider the resulting random fluctuations in the interaction by simply taking a random $\Vkl$. The dependence on both site and particle indices models the random changes expected to be induced by the hopping of any of the concerned particles or of the whole pair. Realistically, there must also be additional dependencies on further neighbors, which are all neglected for simplicity.

Equilibrium states of DPLM are exactly solvable. In particular, particle occupancies $n_i$ follow equilibrium statistics the same as those of a standard identical-particle lattice gas model with a constant interaction energy. These will be explained in \app{sec:eqstats}. Furthermore, equilibrium states of DPLM can be directly constructed using those of standard lattice gas, which exhibits no glassy slowdown (see \app{sec:init}). 

The dynamics of DPLM is defined by standard activated hopping approach for kinetic Monte Carlo simulations.
Specifically, to simulate the dynamics at temperature $T$, each particle can hop to an unoccupied NN site at a rate \cite{fichthorn2000}
\begin{equation}
  \label{w}
    w = w_0 \exp\left(-\frac{E_{0} + \Delta E/2}{k_BT}\right)
\end{equation}
where $\Delta E$ is the change in the system energy due to the hop and $k_B=1$ is the Boltzmann constant. This definition satisfies detailed balance.
We let $E_{0} = 1.5$ so that $E_{0} + \Delta E/2 \ge 0$. Also, we put $w_0 = 10^{6}$ without loss of generality.
Particle motions can be equivalently described as void motions (see \fig{fig:lattice}).
At temperature $T\rightarrow\infty$, DPLM reduces to a simple sliding block model \cite{palmer1990}.


\mainsection{Glassy dynamics}
\label{sec:glassydyanmics}

Let $\phi_v = 1 - \phi$ be the void density where $\phi = N/L^2$ is the particle density in principle related physically to the system pressure. We perform kinetic Monte Carlo simulations of fully equilibrated systems at $L = 100$ at various $T$ and $\phi_v$ (see \app{sec:simulation} for simulation methods).
Standard dynamical measurements show that the system behaves as a simple liquid at high $T$ and $\phi_v$ and a glass at low $T$ and $\phi_v$. As will be further explained, glassy behaviors are shown by the appearance of a plateau in the particle mean square displacement (MSD), a super-Arrhenius $T$ dependence of the particle diffusion coefficient $D$, a stretched exponential form of the self-intermediate scattering function decaying towards zero at long time, a violation of the Stokes-Einstein relation, and typical time and $T$ dependences of a four-point susceptibility. 
In particular, the convergence of the self-intermediate scattering function towards zero rather than a finite value at long time verifies that DPLM is a model of structural glass, as opposed to for example spin glass. For all $T$ and $\phi_v$ studied, DPLM exhibits no sign of ideal glass transition. It also appears ergodic as supported, for example, by the divergence of the particle MSD and the vanishing of the self-intermediate scattering function at long time. 

\subsection{Diffusion coefficient}
\begin{figure}[tb]
\includegraphics[width=\linewidth]{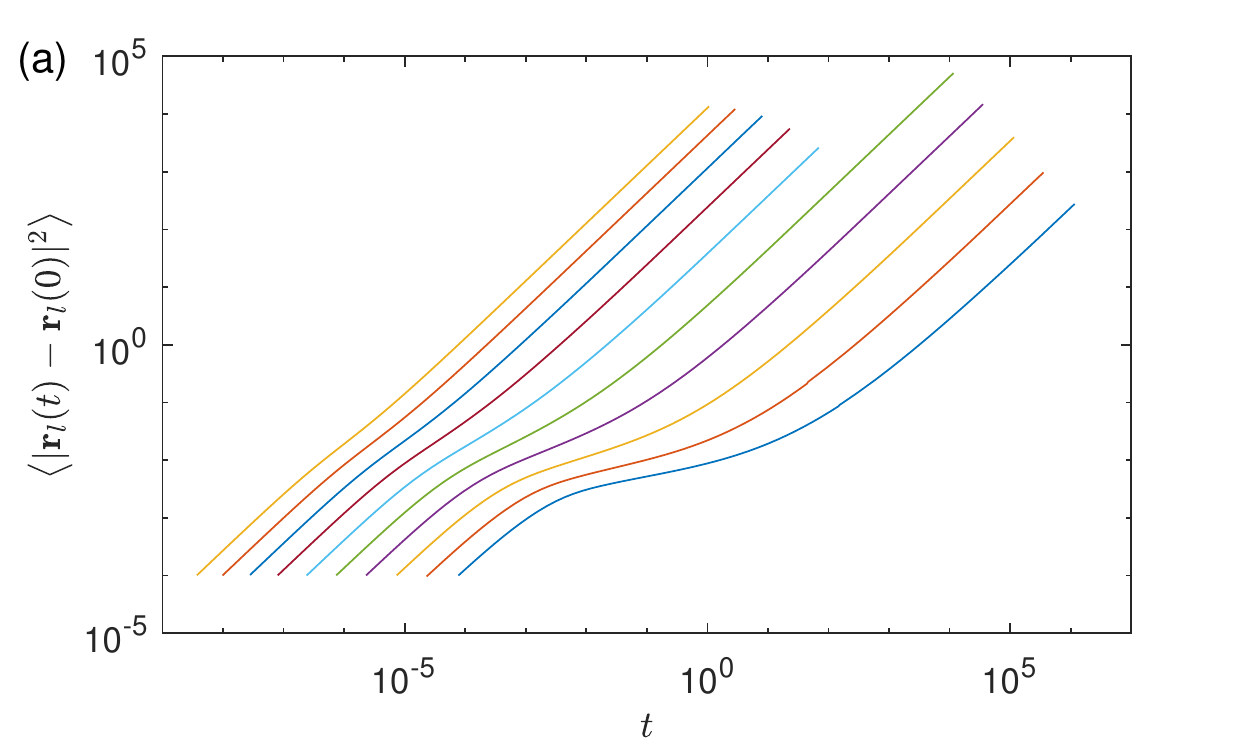}
\includegraphics[width=\linewidth]{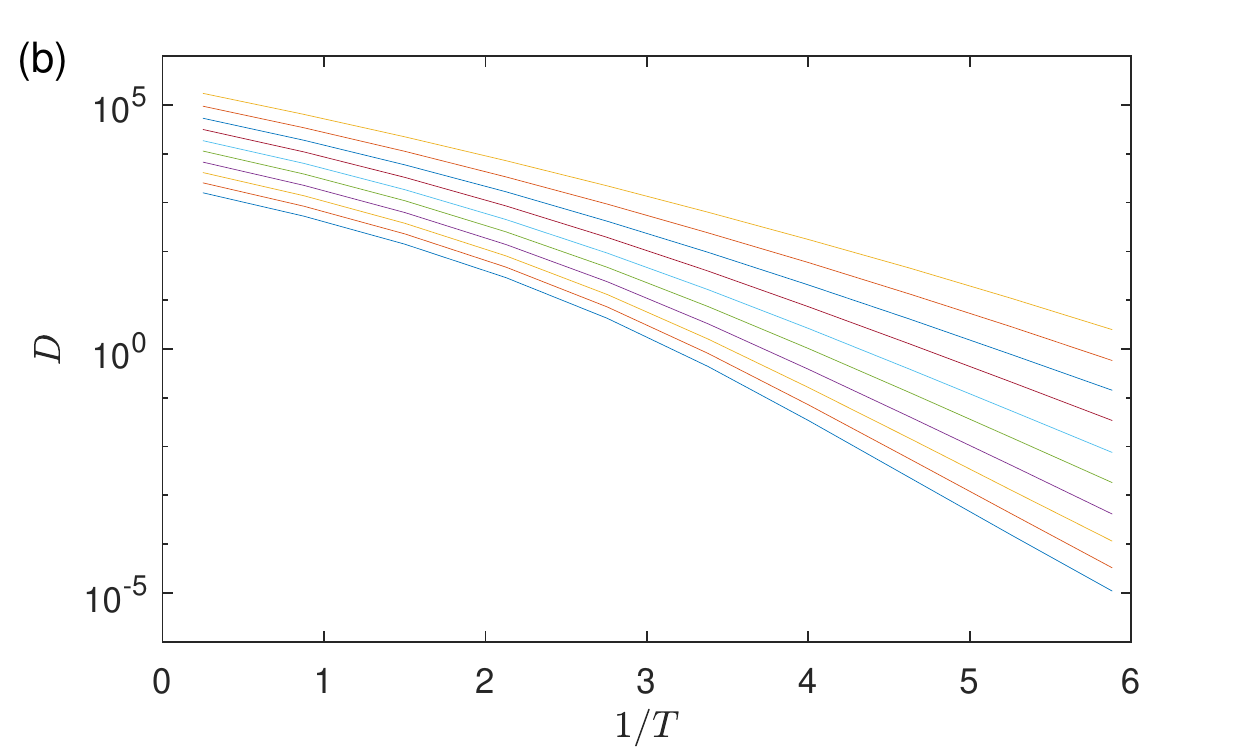}
\caption{
(a) Particle mean square displacement (MSD) against $t$ in log-log scale for $T=$ 0.170, 0.190, 0.216, 0.250, 0.296, 0.363, 0.470, 0.666, 1.142, 4.000 and void density $\phi_v = \phivconst$ with the highest $T$ at the top.
(b) Arrhenius plot of $D$ for $\phi_v=$ 0.005, 0.008, 0.013, 0.021, 0.035, 0.056, 0.092, 0.149, 0.242, 0.392, with the highest $\phi_v$ at the top. 
}
\label{fig:msd}
\end{figure}
We calculate the MSD defined as $\left\langle |\mathbf{r}_l(t) - \mathbf{r}_l(0)|^2 \right\rangle$
where $\mathbf{r}_l(t)$ denotes the lattice position vector of particle $l$ at time $t$.
\Fig{fig:msd}(a) shows the MSD in a log-log plot for different $T$ and $\phi_v=\phivconst$. For $t \to \infty$, the slopes of the lines are consistent with unity, indicating diffusive behavior over long observation time.
Sub-diffusive plateaus appearing at intermediate $t$ at low $T$ indicate cage effects.
Note that being a lattice model without vibrational modes at the sublattice level, the plateaus are much less pronounced as have been found for other lattice models \cite{kob1993}.

From similar MSD for various $T$ and $\phi_v$, we measure the particle diffusion coefficient
\begin{equation}
    D = \frac{1}{2d} \lim_{t \to\infty} \frac{\left\langle |\mathbf{r}_l(t) - \mathbf{r}_l(0)|^2 \right\rangle}{t}
\end{equation}
by fitting to data points where $\left\langle |\mathbf{r}_l(t) - \mathbf{r}_l(0)|^2 \right\rangle > 1$ and the slope in the log-log plot is higher than 0.96.
\fig{fig:msd}(b) shows $D$ in an Arrhenius plot for various $\phi_v$.
It exhibits super-Arrhenius behavior which becomes more pronounced at small $\phi_v$ and low $T$. This shows that DPLM is a fragile glass.

\subsection{Self-intermediate scattering function}
\begin{figure}[tb]
\includegraphics[width=\linewidth]{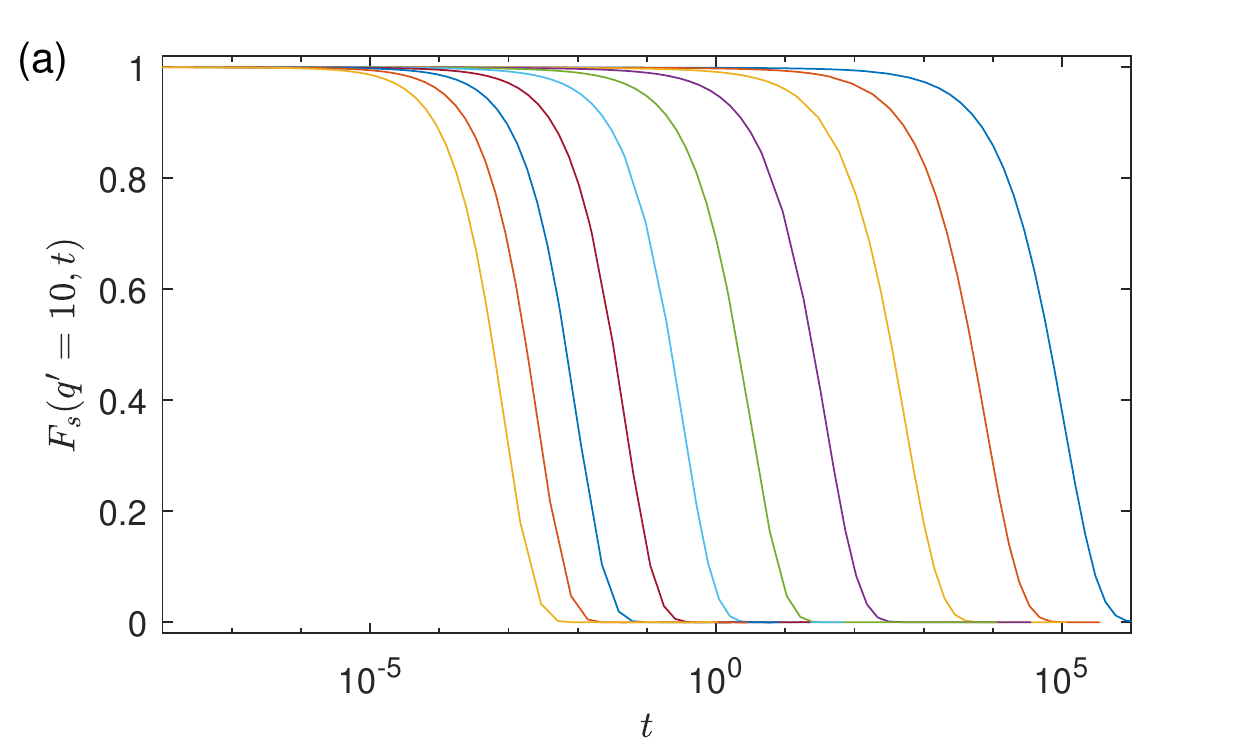}
\includegraphics[width=\linewidth]{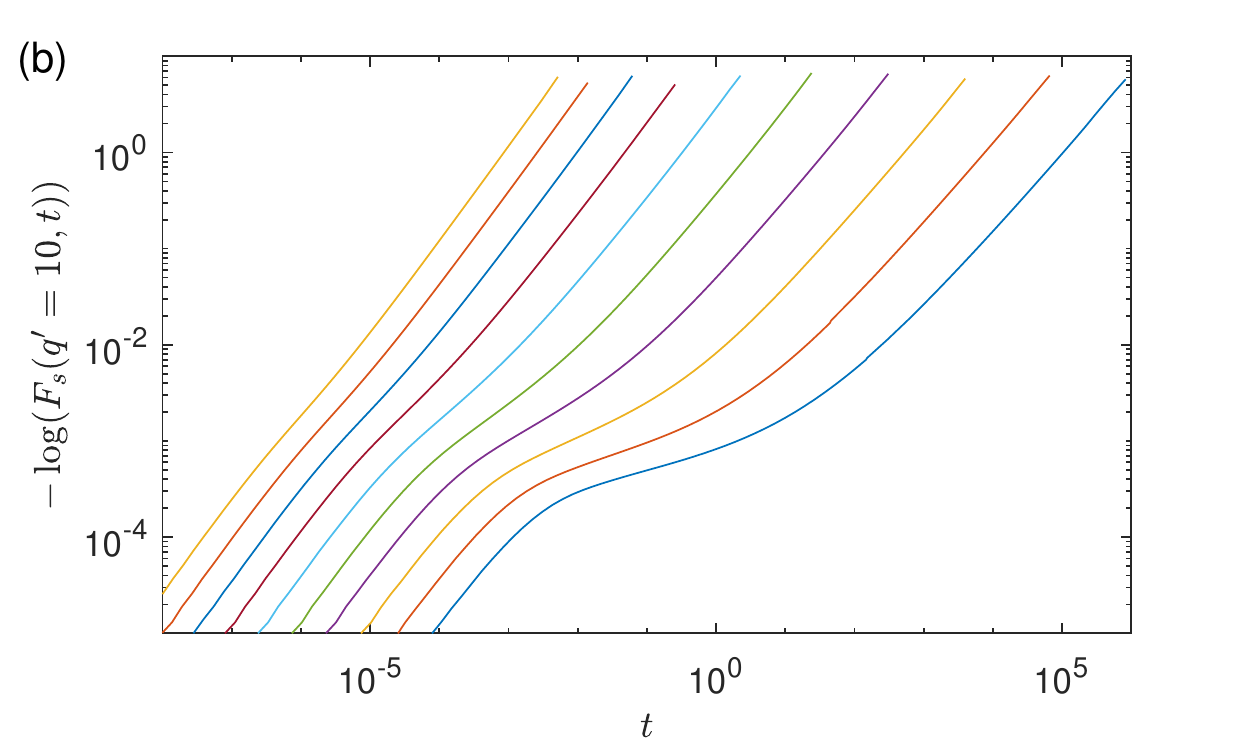}
\caption{
(a) Decays of self-intermediate scattering function $F_s(q,t)$ in linear-log scale, for the same values of $T$ used in \fig{fig:msd}(a), with $T$ decreasing from left to right. Wavenumber $q = (2\pi/L)q' = \pi/5$ and $\phi_v = \phivconst$ are used here.
(b) Same data as in (a) in log-log-versus-log scale. Data corresponding to $F_s(q,t) < 10^{-3}$ are noisy and are omitted.
The slope of the linear region at large $t$ with $10^{-3} \leq F_s(q,t) \leq 0.9$ gives the stretching exponent $\Beta$.
}
\label{fig:sisf1}
\end{figure}
\begin{figure}[tb]
\includegraphics[width=\linewidth]{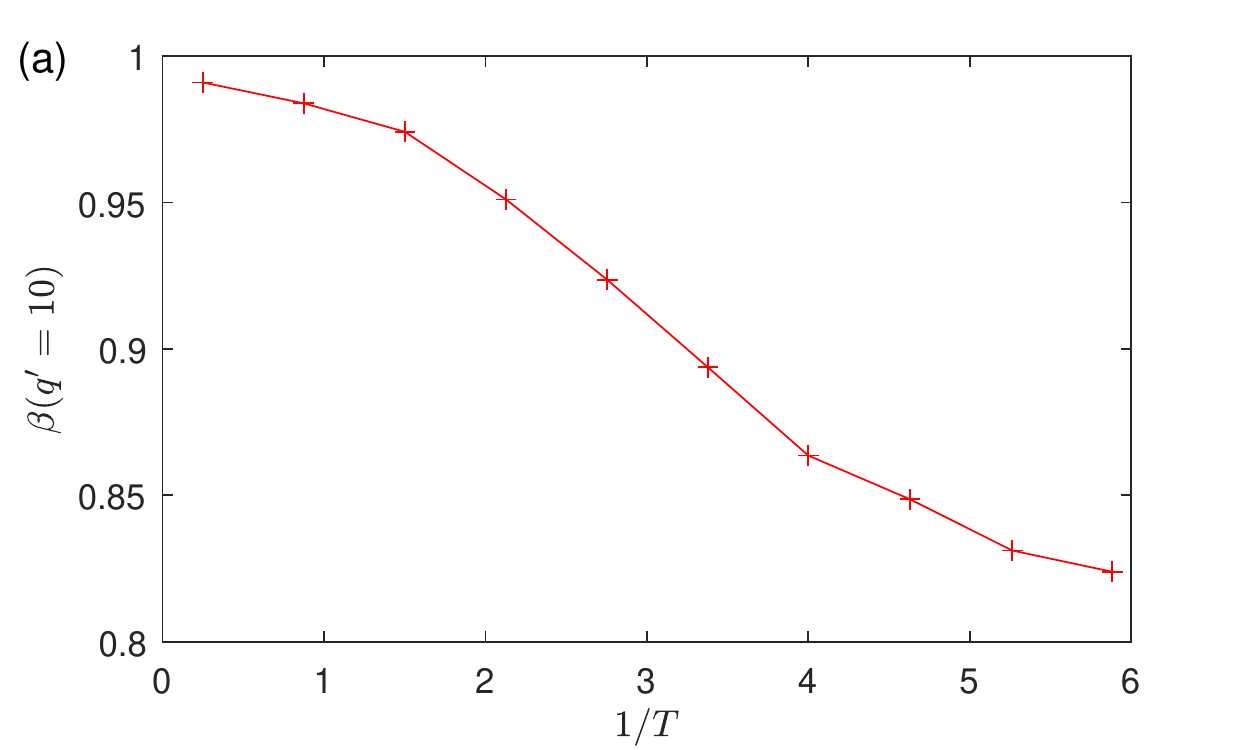}
\includegraphics[width=\linewidth]{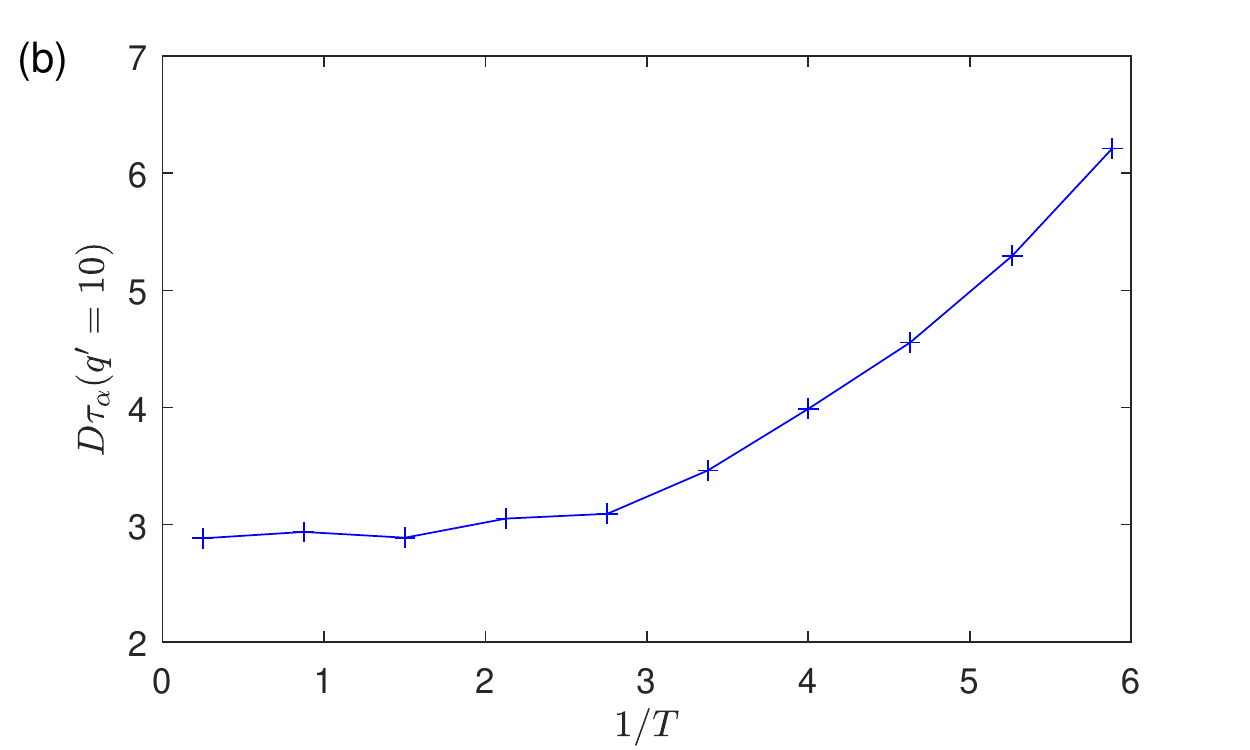}
\caption{
(a) Stretching exponent $\Beta$ plotted against $1/T$ for $\phi_v=\phivconst$.
(b) Violation of the Stokes-Einstein relation, $D\tau_{\alpha} = \text{constant}$ where $\tau_{\alpha}$ is a relaxation time.
}
\label{fig:sisf2}
\end{figure}

We have measured the self-intermediate scattering function defined as
\begin{equation}
    F_s(\mathbf{q},t) = \left\langle e^{i \mathbf{q} \cdot \left( \mathbf{r}_l(t) - \mathbf{r}_l(0) \right)} \right\rangle
\end{equation}
and the result is shown in \fig{fig:sisf1}(a) for $\phi_v=\phivconst$ and $q=(2\pi/L)q' $ with $q'=10$.
A one-step drop of $F_s(q,t)$ versus $t$ instead of a two-step decay is again typical for lattice models \cite{harrowell1993,kob1993,darst2010}.
In glassy systems, the terminal decay of the scattering function is usually well approximated by the Kohlrausch-Williams-Watts (KWW) stretched exponential function of the form $A \exp\left( -(t/\tau)^\Beta \right) $, where $\tau$ is a relaxation time and $\Beta$ ($0 < \Beta < 1$) is the stretching exponent. Our results fit well to the KWW form for large $t$. This is also demonstrated by the log-log plot of $-\log(F_s(q,t))$ against $t$ in \fig{fig:sisf1}(b) which shows a linear region at large $t$ expected from the KWW form with $A\simeq 1$. The stretching exponent obtained from the slope of the linear region is plotted in \fig{fig:sisf2}(a). As $T$ decreases, $\Beta$ drops from 1 to around 0.82, indicating glassy dynamics at low $T$.

From \fig{fig:sisf1}(a), we also extract a relaxation time $\tau_\alpha$ which is the time at which $F_s(q,t)=1/e$. \fig{fig:sisf2}(b) plots $D \tau_\alpha$ against $1/T$. The value clearly increases with decreasing $T$ and demonstrate a violation of the Stokes-Einstein relation expected for glasses.

\subsection{Four-point correlation function}
Close to the glass transition, one region in a glassy fluid can relax much faster than another one. This spatially inhomogeneous dynamical behavior is known as dynamic heterogeneity. To quantitatively study the heterogeneity in the persistence of the particle configuration, one can define an overlap function as
\begin{equation}
    c_l(t,0) = e^{i \mathbf{q} \cdot \left(
        \mathbf{r}_l(t) - \mathbf{r}_l(0)
    \right)}.
\end{equation}
It measures how much particle $l$ moves during times $0$ and $t$ at a length scale $2\pi/q$. Note that the average overlap equals the self-intermediate scattering function $F_s(\mathbf{q}, t)$. Each particle contributes to an overlap field defined by
\begin{equation}
    c(\mathbf{r};t,0) = \sum_l c_l(t,0) \delta\left( \mathbf{r} - \mathbf{r}_l(0) \right)
\end{equation}
Consider its spatial correlation
\begin{equation}
    G_4(\mathbf{r},t)
    = \left\langle  c(\mathbf{r};t,0) c(\mathbf{0};t,0) \right\rangle -
      \left\langle c(\mathbf{0};t,0) \right\rangle^2
\end{equation}
where the average is over the spatial origin $\mathbf{0}$ and the starting time 0. $G_4$ measures the correlation of the fluctuations in the overlap function
between two points that are separated by $\mathbf{r}$.

In the Fourier space, we get
\begin{eqnarray}
    S_4(\mathbf{\tilde q},t)
    &=&  \int e^{i \mathbf{\tilde q} \cdot \mathbf{r}} G_4(\mathbf{r},t) d\mathbf{r} \\
    &=&  N \left\langle \left|
            \frac{1}{N}
            \sum_l e^{i \mathbf{\tilde q} \cdot \mathbf{r}_l(0)}
            \left(
                c_l(t,0) - F_s(\mathbf{q},t)
            \right)
        \right|^2 \right\rangle\nonumber\\
\end{eqnarray}
One can define the susceptibility as $\chi_4(t) = \lim_{\tilde q \to 0} S_4(\mathbf{\tilde q},t)$, which is simply the variance of the overlap function. $\chi_4(t)$ can be interpreted as the typical size of correlated clusters in structural relaxation, thus an efficient measure of the degree of dynamic heterogeneity.

\begin{figure}[tb]
    \includegraphics[width=\linewidth]{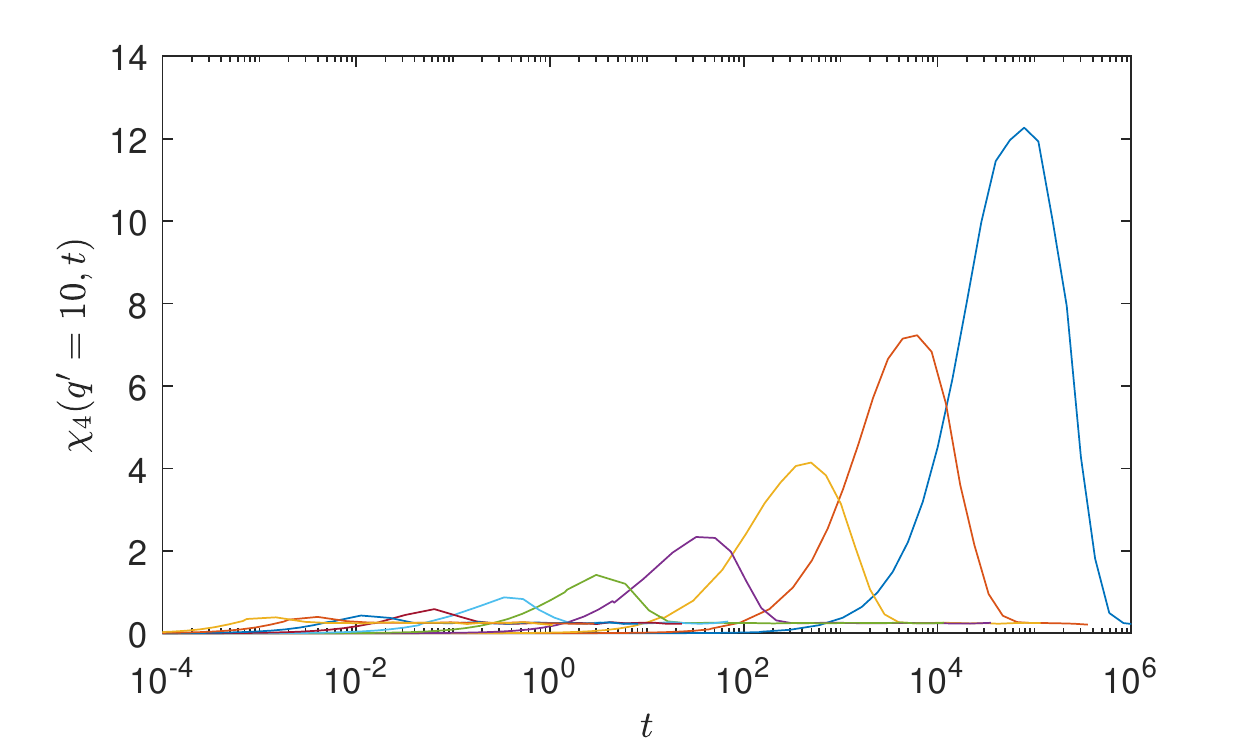}
    \caption{$\chi_4(t)$ for $\phi_v = \phivconst$ and the same values of $T$ used in \fig{fig:msd}(a). $T$ decreases from left to right.}
    \label{fig:chi4}
\end{figure}

\fig{fig:chi4} shows $\chi_4(t)$ from DPLM simulations. As is typical for structural glasses, for each temperature, $\chi_4(t)$ has a peak, which shifts to larger times, and has a larger value when $T$ decreases. This reveals an increasing length scale of dynamic heterogeneity when the system cools down.

\mainsection{Emergent facilitation behaviors}

\begin{figure}[tb]
\includegraphics[width=\linewidth]{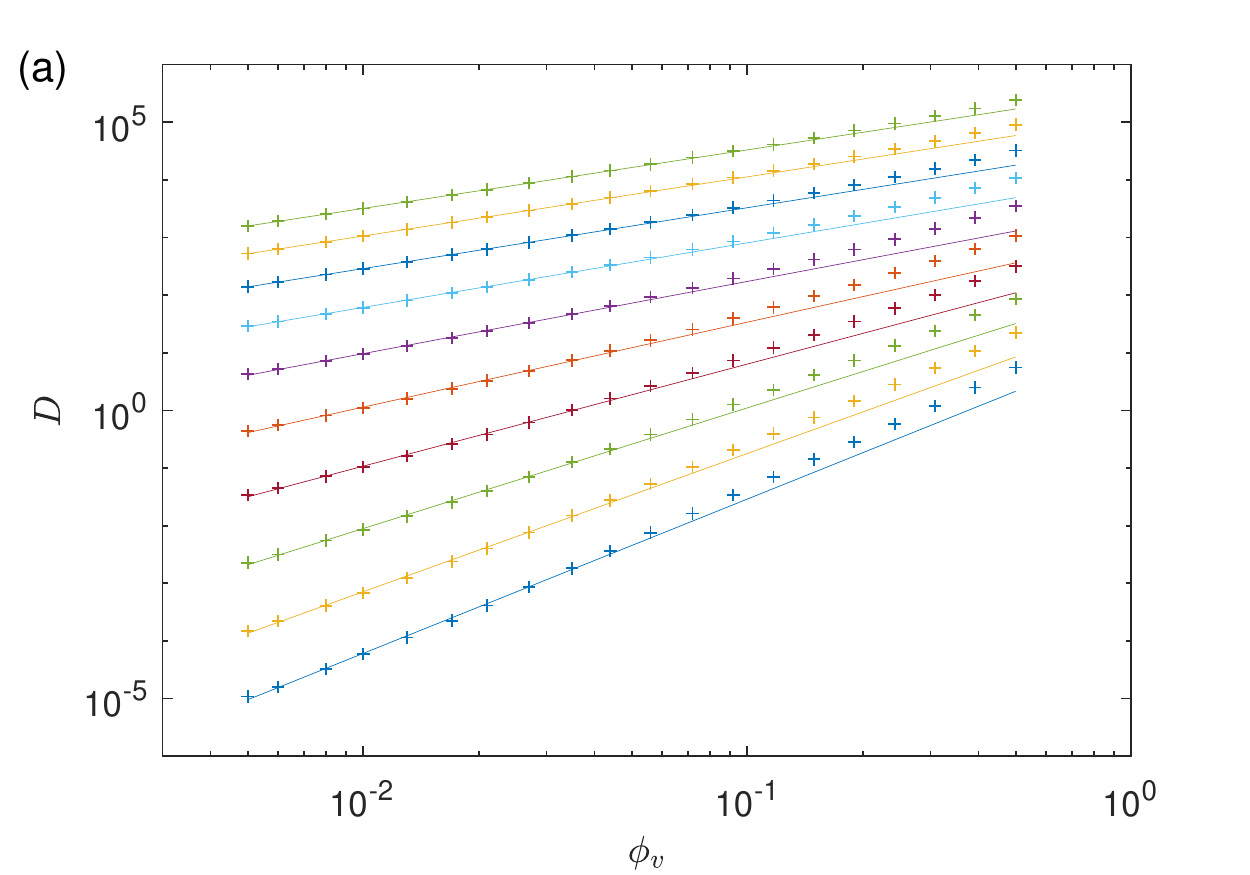}
\includegraphics[width=\linewidth]{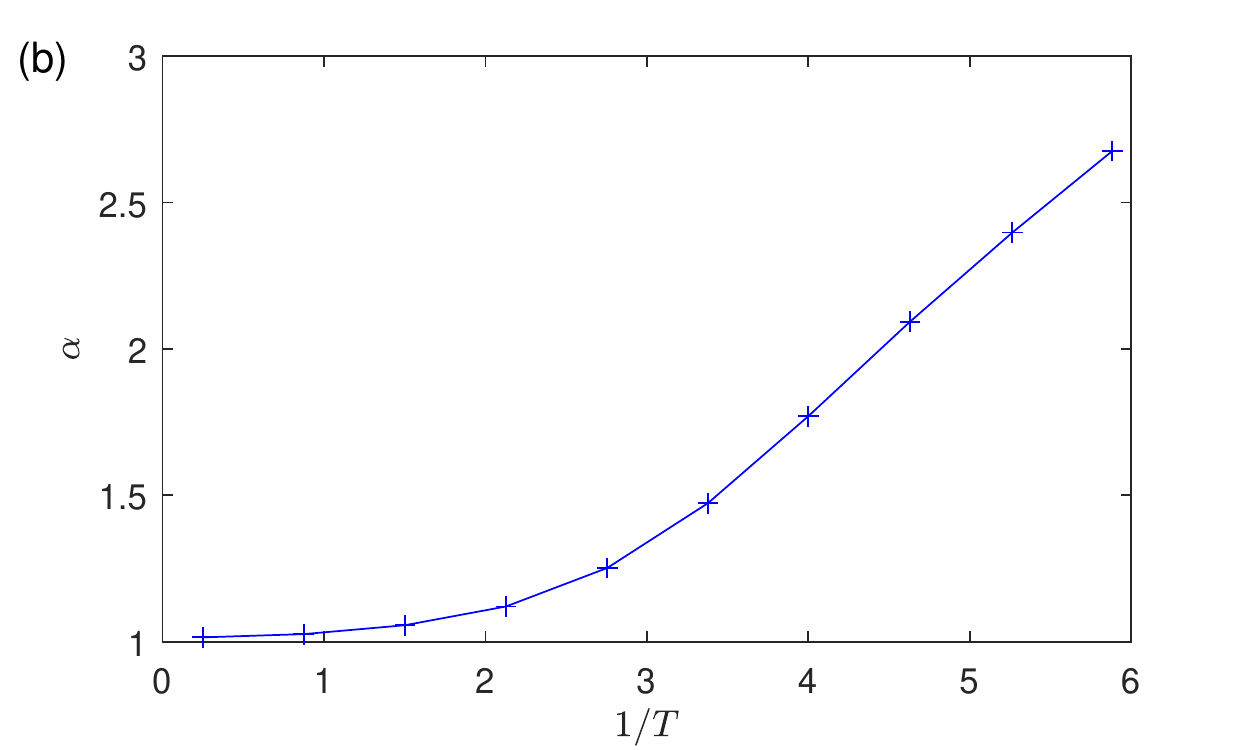}
\caption{
(a) Particle diffusion coefficient $D$ against void density $\phi_v$ in log-log scale for values of $T$ used in \fig{fig:msd}(a) with the highest $T$ at the top. 
(b) Scaling exponent $\alpha$ against $1/T$ obtained from linear fits to data in (a) with $\phi_v \leq 0.05$.
}
\label{fig:facilitation}
\end{figure}

\begin{figure}[tb]
\includegraphics[width=0.8\linewidth]{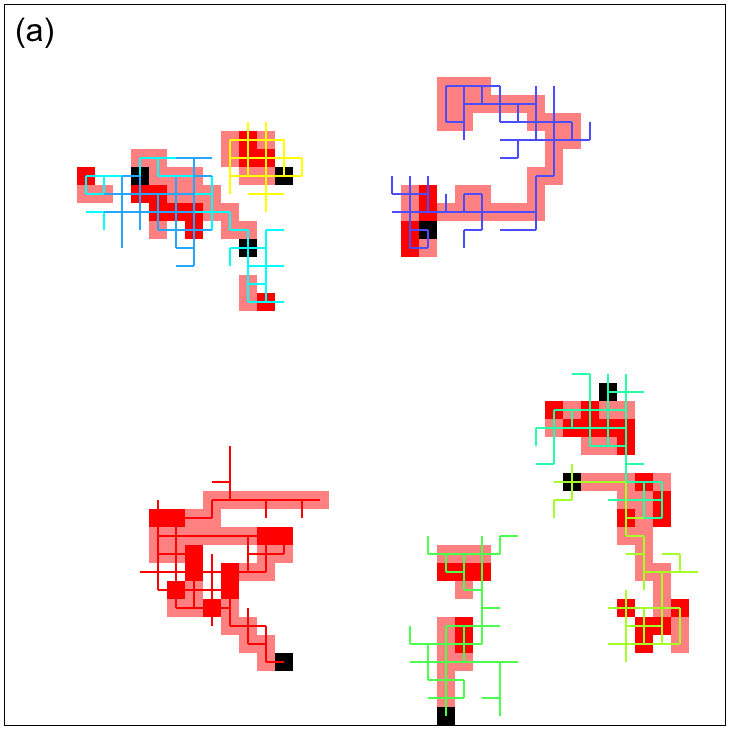}
\includegraphics[width=0.8\linewidth]{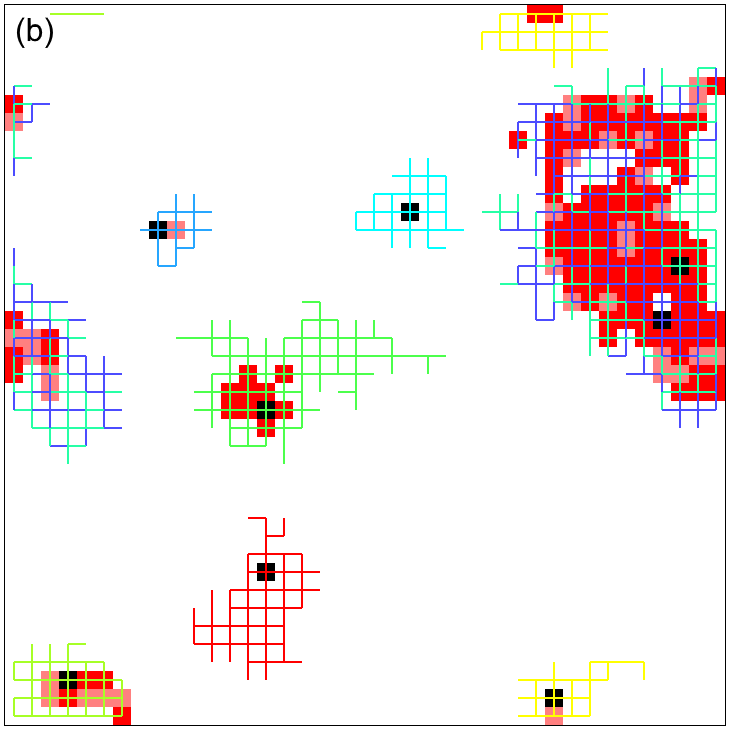}
\caption{ (a) A snapshot from a small-scale simulation on a $40\times 40$ lattice with $1592$ particles and 8 voids, i.e. $\phiv=0.005$. It shows the final positions of voids (black squares) after a short simulation duration of $\Delta \tau=10^{-3}$ at $T=0.5$. Particles with net displacements 0, 1, and $>1$ during the period are shaded in white, pink and red respectively. Each thin line shows the trajectory of a void and is colored randomly. (b) Similar to (a) with $T=0.16$ and $\Delta \tau=5\times 10^{4}$. In both (a) and (b), the particle MSD during the period is about 0.5. Particle dynamics are shown in videos in the Supplemental Material \cite{voidmovie}. }
\label{fig:traj}
\end{figure}

Being an energetically non-trivial model with $T$ and $\phi_v$ independently and fully tunable,
it exhibits much richer physics than purely kinetic models such as KCM.
The particle diffusion coefficient $D$ shown in Fig \ref{fig:msd}(b) is  replotted in \fig{fig:facilitation}(a) against $\phi_v$.
At each $T$, the linear relation in the log-log plot at small $\phi_v$ suggests the power-law
\begin{equation}
  \label{D}
    D \sim \phi_v^\alpha.
\end{equation}
\fig{fig:facilitation}(b) plots the scaling exponent $\alpha$ as a function of $T$. For the liquid state at high $T$, we get $\alpha \simeq 1$ indicating that each void moves independently \cite{palmer1990}. This is supported by a video in the Supplemental Material \cite{voidmovie} showing the motions of the voids as well as the particles at $T=0.5$. It can be observed that voids diffuse independently. \Fig{fig:traj}(a) visualizes the same motions using void trajectories (thin lines). They appear slightly more compact than those of simple random walks due to the disorder. Particles with non-zero net displacements (pink and red) induced by the same void can be grouped into a cluster. Cluster sizes for different voids are relatively uniform. Voids are not trapped and travel throughout the whole system independently at longer times. Dynamic heterogeneity revealed via these clusters is weak.


We now explain that the low $T$ regime exhibits dynamic facilitation \cite{chandler2010review}. \fig{fig:facilitation}(b) shows that $\alpha$ rises to 2 and beyond at low $T$. The nonlinear scaling dictates that a void at small $\phi_v$ has arbitrarily small contributions to the dynamics. According to simple chemical kinetics, $\alpha \simeq 2$ corresponds to motion dominated by pairs of coupled voids.
This quantitatively shows an emergent dynamic facilitation behavior of void motions. It is analogous to KCM and in particular the spin facilitation dynamics of the FA model \cite{fredrickson1984}. We have checked that the nonlinear scaling in \eq{D} is not due to any void aggregation and is robust upon tuning the void-void attraction by a shift of the probability distribution $g$  on the energy scale.

To verify the above facilitation interpretation of \eq{D}, we directly visualize the particle motions for $T=0.16$ in a video in the Supplemental Material \cite{voidmovie}. 
It can be seen that isolated voids are trapped. In sharp contrast, a pair of voids nearby to each other moves vigorously.
\Fig{fig:traj}(b) shows the void trajectories in the same simulation which become very compact with numerous dead-ends indicating confined motions of the voids due to enhanced disorder. The trajectory of each isolated void induces no or few displaced particles (pink and red) as most particles have not hopped or have returned to their original positions. In contrast, the pair of voids nearby to each other induces significantly more extensive intertwining trajectories and vigorous particle displacements. Such pairs dominate the dynamics for the $\alpha \simeq 2$ regime. At longer times, isolated voids typically remain trapped locally by the disorder unless visited and untrapped by other mobile pairs. Pairs of voids may split and new pairs may emerge but these occur at a longer time scale. Dynamic heterogeneity induced by highly mobile pairs of voids among trapped isolated voids is thus strong.
\fig{fig:facilitation}(b) suggests that $\alpha$ may reach 3 and beyond at even lower $T$ indicating dynamics dominated by triplets of voids, etc.

\begin{figure}[tb]
\includegraphics[width=\linewidth]{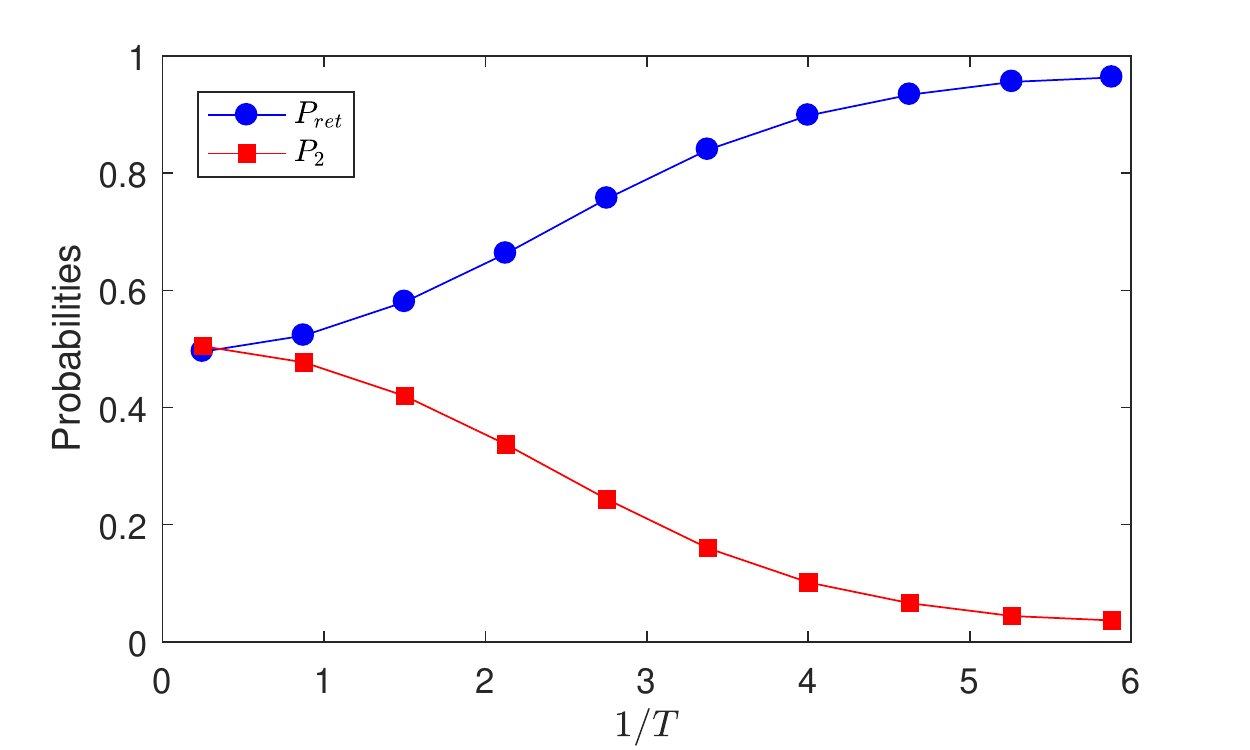}
\caption{Probabilities $P_{ret}$ and $P_2$ for returning and non-returning second hops against $1/T$ for $\phi_v = \phivconst$.
}
\label{fig:pret}
\end{figure}

The dynamics of an isolated void at low $T$ typically involve motions confined along low-energy paths.
Note that $n$ hops by a single void typically corresponds to $n$ single-hops by $n$ particles as shown in \fig{fig:lattice}. A trapped void hence typically leads to bistable-like back-and-forth hops by a few particles.
Such repetitive motions observed in MD simulations of polymers have been argued as the main cause of super-Arrhenius slow-down \cite{lam2015}.
We have adapted the method in Ref. \onlinecite{lam2015} to quantify these repetitions. Specifically, after a particle has hopped, we measure the probability $P_{ret}$ that its next hop returns itself to the original site. The probability $P_{2}$ that it next hops instead to a new site is also measured. The results for $\phi_v=0.01$ are plotted in \fig{fig:pret}. They follow $P_{ret} + P_{2} = 1$ within 0.01\% and the minor deviations are due to particles without a second hop during the observed period. At large $T$, we find empirically that $P_{ret} \simeq 1/2$ applicable for small $\phi_v$ noting that the random walks of voids induce correlated walks of particles \cite{brummelhuis1988}.
We have checked that $P_{ret}$ approaches towards the particle random walk value 1/4 at large $\phi_v$.
As $T$ decreases, $P_{ret}$ increases monotonically reaching 0.96 for the lowest $T$ studied. The trend strikingly resembles those from polymer simulations \cite{lam2015}. This resemblance also strongly supports the physical relevance of DPLM. Such a high $P_{ret}$ means that most hops are reversed and irrelevant to long-time dynamics. The repetition thus must contribute significantly to the slowdown. As $T\rightarrow 0$, our results support $P_{ret}\rightarrow 1$. Most hopping particles then form two-level systems (TLS) known to be relevant to glass at very low $T$ \cite{phillips1987}.

\mainsection{Conclusion}
We have developed DPLM as a lattice gas model based on distinguishable particles for studying glassy dynamics.
In the glassy phase, the particle diffusion coefficient scales nonlinearly with the void density in the low void density limit. This implies that isolated voids are essentially trapped and the dynamics of a void is dominated by facilitation by other voids nearby. Particle hopping becomes increasingly repetitive at low temperature.

DPLM is defined by a simple, generic, and physically motivated system energy function. It has both non-trivial energetics and kinetics. It can be efficiently simulated and equilibrium states can be directly generated at arbitrary temperature and density. Its glassy state does not rely on frustration on a specific lattice type.
These may render DPLM a unique prototypical model for the further study of glassy dynamics and aging in disordered systems.

The definition of DPLM involves no explicit facilitation rule but facilitation behaviors are observed. It thus provides a strong microscopic support to dynamic facilitation and KCM. It will be interesting to deduce the precise coarse-grained lattice model for DPLM. Dynamic facilitation of voids demonstrated by DPLM is analogous to the picture of facilitation via pair-interactions of string-like particle motions motivated by MD simulations of polymers \cite{lam2015}. In that picture, each string is initiated by a single void leading to a one-one correspondence between strings and voids. From \fig{fig:lattice}, the motion of a void alters the particle pairings and hence the energy landscape along its entire path. The energy landscape experienced by another void nearby is thus altered. Whether the second void can diffuse across the path of the first void is thus randomly affected. This demonstrates a form of path interaction of voids which is essentially equivalent to string interactions observed in MD \cite{lam2015}. The particle and void dynamics in DPLM as well as in polymer simulations is recently described on the same footing by a random local configuration tree theory \cite{lam2016}. Alternatively, it will also be of interest to study DPLM defined on the Bethe lattice which may allow exact analysis.

In DPLM, each $\Vkl$ is an independent random variable. More generally, \eq{E} features a very generic Hamiltonian. Adopting instead  a constant $\Vkl \equiv V$ gives a simple interacting lattice gas. As lattice gas models can be mapped to spin models with spin-exchange (Kawasaki) dynamics, it also represents a ferromagnetic or anti-ferromagnetic spin model. Alternatively, a particle-dependent $\Vkl \equiv V_{kl}$ reduces it to a multi-species lattice gas such as a binary alloy \cite{binder1974}.  Limiting to a site-dependent $\Vkl \equiv V_{ij}$, it becomes a variant of the Edwards-Anderson (EA) model for spin glass \cite{edwards1975} with Kawasaki dynamics and a random field. In addition, by continuously varying the correlations between the various $\Vkl$, \eq{E} describes models interpolating between these systems. 

\acknowledgements
We thank helpful discussions with J.Q. You,  Ho-Kei Chan and Peter Harrowell. We are grateful to the support of Hong Kong GRF (Grant 15301014).

\appendix

\section{Exact equilibrium statistics}
\label{sec:eqstats}
Assuming ergodicity, which is supported by our simulations in Sec. \ref{sec:glassydyanmics}, it is possible to derive exact equilibrium states of DPLM in the thermodynamic limit. This is because the system follows a Boltzmann distribution which factorizes over the bonds. More specifically,
equilibrium statistics in the ergodic phase of a system with $N$ particles is described by the canonical partition function 
\begin{equation}
    \label{Z}
    Z =  \sum_{\s} e^{-\upbeta E}
\end{equation}
where the sum is over all possible system states $\s$ and $\upbeta=1/k_BT$.
Noting that $s_i=0$ denotes a void, $Z$ can be rewritten as
\begin{equation}
    \label{Z2}
    Z = \sum_{\n} \sumsiN e^{-\upbeta E} .
\end{equation}
Here, the first sum is over all possible site occupancies  $\n$ with
 $n_i$ defined in \eq{n}. The second sum is over the set $\mathcal{P}_N$ of the $N!$ permutations of particle arrangement $\sP$ at the $N$ occupied sites with $n_i=1$.
\Eq{Z2} can be recast into
\begin{eqnarray}
  \label{Z3}
    Z = \sum_\n \Zn 
\end{eqnarray}
where $\Zn$ is the partition function restricted to the specific site occupation $\n$ and is given by
\begin{eqnarray}
  \label{Zn1}
    \Zn = \sumsiN \prod_{<ij>'}  e^{-\upbeta \Vij}  ~.
\end{eqnarray}
after applying \eq{E2}. 

\subsection{Quenched and annealed averaging} 
\label{sec:quenched}
The value of primary interest is the quenched average $\avq{\ln Z}$ where the bar denotes averaging over the time-independent variables $\Vkl$. 
At sufficiently high $T$, it may agree with the annealed average $\ln \ava{Z}$, where $\ava{\cdot}$ again denotes averaging over $\Vkl$ which is now reinterpreted as additional time-dependent system state variables. A detailed derivation of this agreement will be explained in \app{sec:perm}.

We first study annealed averages which are much easier to calculate. Applying annealed averaging to \eq{Zn1}, we get
\begin{eqnarray}
\label{Zna}
    \avaZn
    &=& \sumsiN \prod_{<ij>'} \ava{ e^{-\upbeta \Vij} }
\end{eqnarray}
where we have noted that each $\Vij$ has a distinct set of indices and are thus independent random numbers.
Defining
\begin{eqnarray}
\label{eU}
e^{-\upbeta U} &=&  \ava{ e^{-\upbeta\Vij} },
\end{eqnarray}
$U$ can then  be interpreted as the average free energy of a bond between two NN particles and is given by
\begin{eqnarray}
\label{U}
U  &=& -\frac{1}{\upbeta} \ln \int_{-\infty}^\infty { e^{-\upbeta V} g(V) dV }.
\end{eqnarray}
Substituting \eq{eU} into \eq{Zna}, all terms in the sum become identical and this trivially gives
\begin{eqnarray}
    \avaZn
    &=& N! \prod_{<ij>'} e^{-\upbeta U} .
\end{eqnarray}
It further reduces to 
\begin{eqnarray}
    \label{aZn}
    \avaZn = N! ~ e^{-\upbeta N_b U}
\end{eqnarray}
where $N_b$ is the number of pairs of bonded particles for the given site occupation $\n$. Substituting into the annealed average of \eq{Z3}, we get
\begin{eqnarray}
    {\ava Z} = {N!}
     \sum_\n  e^{-\upbeta N_b U}
\end{eqnarray}
The factor $N!$ 
results from the particle distinguishability and is related to the Gibb's paradox \cite{kubo1992book}. It is irrelevant and can be omitted for canonical ensembles with a constant $N$ considered here. We thus redefine $Z$ by multiplying with a factor $1/N!$ and obtain
\begin{eqnarray}
    \label{Za}
    \ava {Z}  =
     \sum_\n  e^{-\upbeta N_b U} .
\end{eqnarray}

\subsection{Averaging over permutations}
\label{sec:perm}
As discussed above, it is valid to redefine $Z$ with an additional factor $1/N!$. Specifically, we continue to adopt \eq{Z3} while \eq{Zn1} is replaced by
\begin{eqnarray}
  \label{Zn2}
    \Zn = \frac{1}{N!} \sumsiN  \prod_{<ij>'}  { e^{-\upbeta \Vij} }  .
\end{eqnarray}
The r.h.s. now involves explicitly an average over particle permutations among the occupied sites. 
We will now explain that this averages out all $\Vij$. This is because as emphasized in \eq{E3}, $\Vij$ with nontrivial indices $s_i$ and $s_j$ is not quenched. It indeed samples over many different $\Vkl$ as particles permute, in sharp contrast to the quenched $\Vkl$ at fixed indices. 

Without loss of generality, assume that sites 1 and 2 are occupied nearest neighboring sites. We single out the permutations concerning sites 1 and 2, giving
\begin{eqnarray}
  \label{Znsplit}
    \Zn &=& \frac{1}{N(N-1)} \sum_{\substack{s_1,s_2=1\\s_1\neq s_2}}^N  e^{-\upbeta V_{12s_1s_2} } Z_{N-2}(s_1,s_2) \nonumber\\ 
\end{eqnarray}
where $Z_{N-2}(s_1,s_2)$ is the partition function for the remain $N-2$ sites  excluding particles $s_1$ and $s_2$ defined as
\begin{eqnarray}
\label{ZN2}
Z_{N-2}(s_1,s_2)
&=&
\frac{1}{(N-2)!} \nonumber\\
\times&&\sum_{ \substack{\{s_i>0\}\in \mathcal{P}_{N-2}~ \\ {s_i \neq s_1,s_2}} }  \prod_{\substack{<ij>'\\\{i,j\}\neq \{1,2\}}}   e^{-\upbeta \Vij}  . ~~~~~~~~~
\end{eqnarray}
Since all particles are statistically equivalent, the dependence of $Z_{N-2}(s_1,s_2)$ on $s_1$ and $s_2$ is a manifestation of random fluctuations resulting from the random $\Vkl$. Assuming negligible fluctuations in  $Z_{N-2}(s_1,s_2)$ at large $N$, which will be justified later,
we write  $Z_{N-2}\equiv Z_{N-2}(s_1,s_2)$ 
and \eq{Znsplit} reduces to
\begin{eqnarray}
  \label{Zn7}
    \Zn &=&  e^{-\upbeta U} Z_{N-2} 
\end{eqnarray}
where $U$ is defined in \eq{U}.
Repeating similar procedures, a factor $e^{-\upbeta U}$ is contributed by every bond and we get
\begin{eqnarray}
    \label{Zn9}
    \Zn =  e^{-\upbeta N_b U}
\end{eqnarray}
analogous to  \eq{aZn} after irrelevant prefactors in the latter are dropped.

At finite $N$, $\Zn$ for a given set of $\Vkl$ deviates from the value in \eq{Zn9} with a magnitude characterized by the standard deviation $\sigma_Z$. We will show that $\sigma_Z$ becomes negligible compared with $\Zn$ as $N$ increases.
At high $T$, this is obvious because fluctuations of each factor $e^{-\upbeta \Vij}$ in \eq{Zn2} is small. We thus focus only on the case of low $T$. The deduction is non-trivial because terms in \eq{Zn2} are correlated and have   large variances  increasing with $N$.

At low $T$, a term in \eq{Zn2} is significant predominantly when all its   factors are relatively large. We thus characterize each factor only by whether it is large or small via the approximation 
\begin{eqnarray}
\label{appro}
\expV \simeq \frac{\fij}{2k_BT} e^{-\upbeta U} 
\end{eqnarray}
where
\begin{eqnarray}
\label{xi}
\fij =
\begin{cases}
1  & \text{for $\Vij \in [V_0, V_0+2k_BT] $}\\
0 &  \text{otherwise}
\end{cases}
\end{eqnarray}
with $V_0=-0.5$.
Noting that the apriori probability density $g(\Vij)$ of $\Vij$ is uniform in $[V_0, V_0+1]$, we have constructed the approximation so that the average $e^{-\upbeta U}$ of $\expV$ is  unchanged. In particular, 
the probability $p$ that $\fij=1$ is 
\begin{eqnarray}
\label{p}
p = 2 k_BT.
\end{eqnarray}
\Eq{Zn2} is then approximated by 
\begin{eqnarray}
\label{M0}
\Zn \simeq \frac{1}{N!} ~ p^{-N_b} e^{-\upbeta N_b U} ~M 
\end{eqnarray}
where 
\begin{eqnarray}
  \label{M}
    M =  \sumsiN  \prod_{<ij>'} \fij .
\end{eqnarray}
Here, $M$ equals the number of relevant particle permutations which contribute significantly to $\Zn$. For each of these permutations, it is easy to see that all interactions are within $k_BT$ from the average value $V_0+k_BT$.

We now evaluate the statistical properties of $M$ by tackling the combinatorial problem of counting the relevant permutations.
For simplicity, we illustrate further calculations for a fully occupied $N \times 1$ lattice with interactions only in the non-trivial dimension, but generalization is  straightforward.
First, there are $N$ ways to occupy site $1$. For each choice, there are on average $(N-1)p$ ways to occupy site $2$ in which $\xi_{12}=1$. It is analogous for the other sites except for $i=N$ which contributes a factor $p^2$ because both $\xi_{N-1,N}$ and $\xi_{1N}$ must be nonzero. The average of $M$ is thus
\begin{eqnarray}
  \label{barM}
  \barM &=& N \cdot (N-1)p \cdot (N-2)p \cdots  1p^2 \nonumber\\
  &=& N! ~p^N 
\end{eqnarray}
As a consistency check, substituting \eq{barM} into \eq{M0} and assuming $M\simeq \barM$ recovers \eq{Zn9}.

More importantly, we now calculate the standard deviation $\sigma_M$ of $M$. 
For each of the $N$ ways to occupy site 1, the number of ways to occupy site $2$ follows a binomial distribution with a variance $(N-1)p(1-p)$. Each of these choices at sites 1 and $2$ on average results at $(N-2)!p^{N-1}$ relevant ways to permute the remaining $N-2$ particles. Therefore, fluctuations at $i=2$ contribute a variance $v_2$ to $M$ given by
\begin{eqnarray}
  v_2 &=& N \times (N-1)p(1-p) \times  [(N-2)!p^{N-1}]^2 \nonumber \\
  &=& \frac{(1-p)\barM^2}{N(N-1)p}
\end{eqnarray}
where we have used \eq{barM}. We next consider fluctuation at site 3 as a further example. For each of the on average $N(N-1)p$ ways to occupy sites 1 and 2, the number of ways to occupy site 3 follows a binomial distribution with a variance $(N-2)p(1-p)$. Each of these choices at sites 1, 2 and 3 on average results at $(N-3)!p^{N-2}$ relevant ways to permute the remaining $N-3$ particles. Fluctuations at site $3$ thus contribute a variance $v_3$ to $M$ given by
\begin{eqnarray}
  v_3 &=& N(N-1)p \times (N-2)p(1-p) \times [(N-3)!p^{N-2}]^2 \nonumber \\
  &=& \frac{(1-p)\barM^2}{N(N-1)(N-2)p^2}
\end{eqnarray}
Fluctuations at other sites can be similarly calculated.
Neglecting correlations between these fluctuations, we get $\sigma_M^2=\sum_{i=2}^N v_{i} $ which simplifies to
\begin{eqnarray}
\sigma_M^2 &=& (1-p)\barM^2 \left( \frac{1}{N(N-1)p} + \frac{1}{N(N-1)(N-2)p^2} \right.\nonumber \\
   && \left. + \frac{1}{N(N-1)(N-2)(N-3)p^3} + \cdots ~~ 
\right)
\end{eqnarray}
For large $N$, all but the first term are negligible and  we get 
$\sigma_M \simeq \sqrt{(1-p)/p} ~ {\barM }/{N}$.  
Since $\Zn \propto M$ according to \eq{M0}, the standard deviation of $\Zn$ is $\sigma_Z \simeq \sqrt{(1-p)/p} ~ \Zn/N $. In particular, we have
\begin{eqnarray}
\label{sigmaZ}
\sigma_Z & \sim & \frac{\Zn}{N}  
\end{eqnarray}

    To verify this result, we have numerically performed direct enumeration of $10^5$ values of $\Zn$ using either \eq{Zn2} or \eq{M0} for $10^5$ independent realizations of $\Vkl$ for $N\le 11$ and $T\ge 0.2$. In both cases, \eq{sigmaZ} is readily verified. As a further check of our method, we consider  alternative interactions in the form $\Vkl\equiv V_{kl}$, representing particle-dependent interactions as opposed to site-particle-dependent ones. Using analogous arguments, we find instead $\sigma_Z \sim \Zn$, which is also well verified numerically by direct enumeration.

It is straightforward to generalize \eq{sigmaZ} to arbitrary site occupancies $n_i$ in 2D. Therefore, for DPLM studied in this work, \eq{Zn9} admits corrections only of order $1/N$ and is essentially exact for large $N$. Substituting \eq{Zn9} into \eq{Z3}, we get 
\begin{eqnarray}
    \label{Zfinal}
    {Z} =
     \sum_\n  e^{-\upbeta N_b U} 
\end{eqnarray}
where all $\Vkl$-dependent correction terms are of higher orders in $1/N$.
Note that similar arguments also imply that $Z_{N-2}(s_1,s_2)$ defined in \eq{ZN2} has negligible fluctuations and this justifies the assumption used in deriving \eq{Zn7}.

We emphasize that we have $not$ at this point performed the ensemble average over $\Vkl$ and $Z$ in \eq{Zfinal} have already become independent of $\Vkl$ due to the averaging over particle permutations. This is quite analogous to self-averaging behaviors exhibited by many systems. Here, sample to sample fluctuations of $Z$ hence vanish and all quenched averaging becomes trivial, i.e. $\avq{\ln Z} = \ln Z$.
A further comparison of \eq{Zfinal} with \eq{Za} gives
\begin{eqnarray}
\label{Zequal}
\avq{\ln Z} = \ln Z =  \ln \ava Z.
\end{eqnarray}
This shows the identical statistical properties of quenched and annealed ensembles in the ergodic phase for large $N$. 

\subsection{Equilibrium properties}

Let $Z_{LG}$ be the partition function of a simple identical-particle lattice gas with a NN particle interaction energy $U$. It is easy to see that $Z_{LG}$ is in fact identical to $Z$ in \eq{Zfinal}, i.e. \begin{eqnarray}
  \label{ZLG}
  Z = Z_{LG} .
\end{eqnarray}
Therefore, DPLM and simple lattice gas have exactly the same equilibrium particle occupation statistics despite the very different dynamics. A simple lattice gas has a gas-liquid phase transition at the vaporization temperature $T_v$, which depends on $U$ and thus on the distribution $g$.  The lattice gas  can be further mapped to the 2D Ising model with an exchange $J=-U/4$ \cite{binder1974}. Applying Onsager's solution $T_v = 2J/\ln(1+\sqrt{2})$
for the 2D Ising model \cite{kubo1992book}, we get 
\begin{equation}
\label{Tv}
T_v = \frac{-U}{4\ln(1+\sqrt{2})}
\end{equation}
where $U$ is given in \eq{U} evaluated at $T=T_v$.
Solving Eqs. (\ref{U}) and (\ref{Tv}) numerically, we get $T_v \simeq 0.132$. We have verified this value of $T_v$
using small-scale DPLM simulations at e.g. $\phiv=0.5$. Since $T_v$ is below $T$ studied in our main simulations,
the systems considered here correspond to lattice gases in the gaseous phase, in which particles are only slightly attractive and neither particles nor voids in dilute concentration aggregate.

We now derive the equilibrium distribution of the interactions for annealed ensembles, which is identical to that of quenched ensembles according to \eq{Zequal}. Restricting our consideration to a given site occupancy $\n$. We study the equilibrium properties of the remaining state variables $s_i>0$ and $\Vkl$.  They follow the Boltzmann probability distribution
\begin{equation}
  P_{eq}(\s,\{\Vkl\}) \propto ~ e^{-\upbeta E} \prod_{{<i,j>,k,l}} g(\Vkl)
\end{equation}
where the product is over all NN sites $i$ and $j$ and all particles $k$ and $l$. Applying \eq{E2}, we get
\begin{eqnarray}
  \label{Peq}
  P_{eq}(\s,\{\Vkl\}) \propto 
\left[ \prod_{<i,j>'} e^{-\upbeta \Vij}  g(\Vij) \right]  ~~~~~\nonumber\\
\times \left[ \prod_{\{<i,j>,k,l \} \in  C} g(\Vkl) \right ] ~~~~~~~~~~~~~~
\end{eqnarray}
Here, the first product is restricted to bonded NN sites $i$ and $j$. Thus, the realized interaction $\Vij$ which describes an existing bond in the state $\s$ follows the Boltzmann distribution
\begin{equation}
  \label{peq}
  p_{eq}(\Vij) = \frac{1}{\mathcal N} e^{-\upbeta \Vij}  g(\Vij)
\end{equation}
where $\mathcal N = \int e^{-\upbeta V} g(V) dV$ is a normalization constant.
The second product in \eq {Peq} is over the complementary set $C$ of unrealized interactions $\Vkl$ which do not represent any existing bond in the state $\s$. This equation also implies that these unrealized interactions simply follow $g(\Vkl)$.

We now further derive some other useful results. Adopting annealed ensemble, all $s_i$ in the r.h.s. of \eq{Peq} are dummy indices of identical independent variables. All permutations of $s_i$ are indeed equivalent and only amount to different labeling of the particles. To see this mathematically, we note that after    
integrating $P_{eq}(\s,\{\Vkl\})$ over all $\Vkl$, we get a uniform probability distribution 
\begin{equation}
\label{Peqs}
P_{eq}(\s) = 1/N!
\end{equation}
demonstrating the equivalence of all $N!$ permutations $\s$ for annealed ensembles as expected.

In addition, the average interaction between bonded particles is
\begin{eqnarray}
  \av{\Vij} = \int V p_{eq}(V) ~ dV
\end{eqnarray}
where $p_{eq}$ is given in \eq{peq}. Using \eq{E2}, the average energy per particle is then 
$\av E / N = \av {N_b/N} \av{ \Vij}$. 
For small $\phiv$ with mostly isolated voids, the average number of bonds per particle is 
$\av {N_b/N} \simeq 2 (1-\phiv)$. This gives
\begin{equation}
  \label{Eav}
   \frac{\av E}{N} = 2 (1-\phiv) \int V p_{eq}(V) ~ dV.
\end{equation}

\section{Simulation details}
\label{sec:simulation}

We will describe both elementary and accelerated simulation approaches, which have been checked to generate statistically identical results.  Our main simulations are all performed using accelerated algorithms. Each of them at lattice size $L=100$ takes up to about 20 hours to run on an Intel Xeon processor core. Data for each set of values of $T$ and $\phi_v$ are typically averaged over 5 similar independent runs.  Additional shorter runs recording particle positions at a higher time-resolution are also needed to obtain correlation data at short time.

\subsection{Elementary kinetic Monte Carlo method}
Simulations can be performed using standard kinetic Monte Carlo approach.
At each time step $\Delta t$, the following procedures can be performed:
\begin{itemize}
\item Randomly choose a site $i$.
\item Randomly choose a site $j$ which is a NN of $i$.
\item If $n_i=1$ and $n_j=0$ is false, reject this step.
\item Accept particle hop from $i$ to $j$ with probability $4L^2 w \Delta t $  where $w$ is calculated using \eq{w}
\end{itemize}
Here, $\Delta t$ must be small and satisfies $4L^2 w \Delta t \le 1$ for all possible configurations.

\subsection{Rejection-free method}
The simple kinetic Monte Carlo algorithm above is inefficient due to too many rejected move attempts. A rejection-free method \cite{bortz1975} is much more efficient. Let $N_v = L^2 - N$ be the number of voids. We optimize our algorithm for $\phi_v \simeq 0$  which is most demanding due to the slow dynamics. The number of possible hops is $4N_v$ in general. The associated hopping rates $w$ are calculated using \eq{w} and stored at the lowest level of a complete binary tree. Each parent node then stores the sum of the two immediate children. Note that an exchange of two voids is unphysical and is assigned a rate 0.

For each time step $\Delta t$, one of the $4N_v$ possible hops is randomly selected with a relative probability $w$.
It is straightforward to select the hop efficiently by randomly descending the binary tree using the node values as the relative probabilistic weights. The hop is then executed. A few hopping rates associated with the hopping particle and its neighbors are recalculated since the local configuration has changed. The binary tree is then also updated accordingly.
It is easy to see that $\Delta t$ is time dependent and follows $\Delta t = 1/w_{root}$, where $w_{root}$ is the value at the root of the binary tree and equals the sum of all the $4N_v$ rates  \cite{lam2008}.

\subsection{Two-step interaction energy tabulation}
A nontrivial point in the programming for DPLM is that the total number of $\Vkl$ is of order $N^2L^2 \sim N^3$.  This requires too much memory storage for large $N$. For medium values of $N$, $\Vkl$ can be sampled only when needed and stored using a hash data structure. In our main simulations with a large $N\sim L^2 = 10^4$, it is necessary to adopt a two-step tabulation method to be explained below. 

As an approximate scheme, we put
\begin{equation}
  \Vkl = v(Q_{i}(k),Q_{j}(l)) .
  \label{Vpermute}
\end{equation}
Here, each $Q_{i}$ for site $i$ is an independent random permutation function mapping the set $1,2,\dots, N$ to itself.
The function $v$ thus involves only order $N^2$ tabulated random numbers sampled from $g$ which are independent from each other except when the symmetry $v(k,l)=v(l,k)$ applies. Before simulation starts, the functions $v$ and $Q_{i}$ are randomly sampled and stored.  The memory requirement significantly decreases from order $N^3$ to order $N^2$.
The method do introduce some unwanted correlations between the ideally independent $\Vkl$.
However, we have checked in medium scale simulations that it gives results statistically identical to those using the hash-table method.

\begin{figure}[tb]
\includegraphics[width=\linewidth]{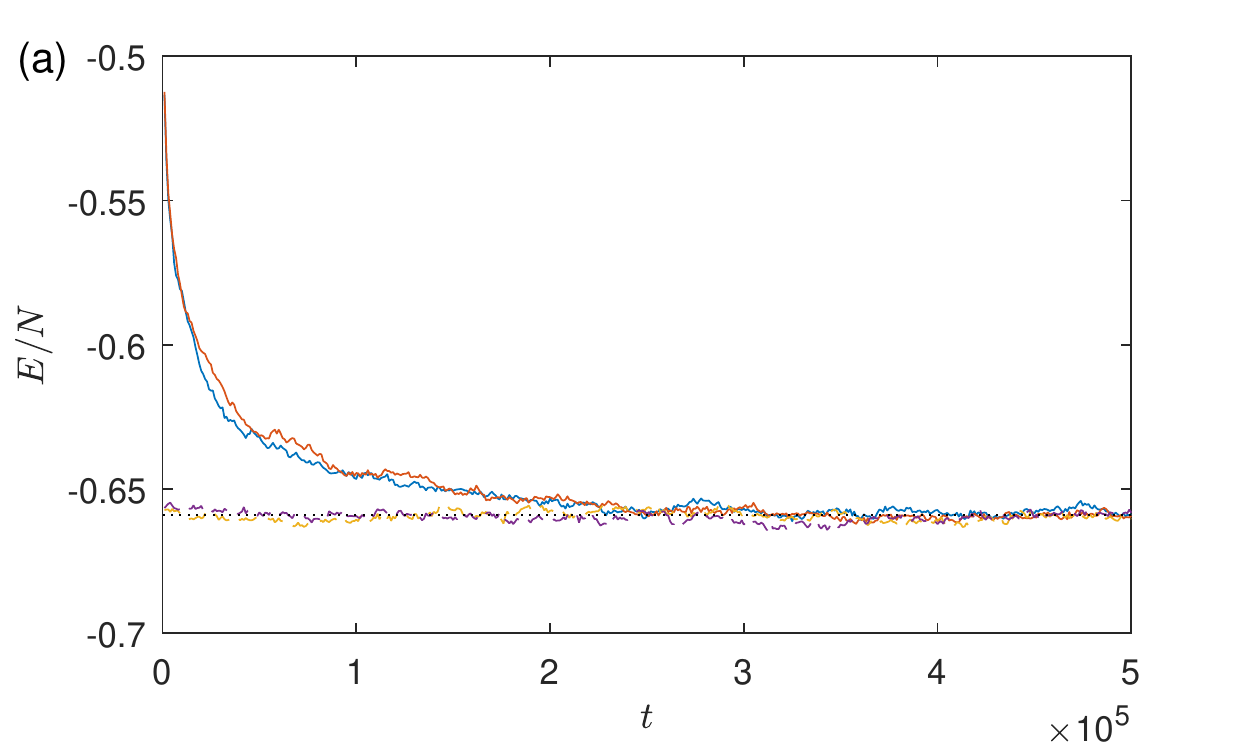}
\includegraphics[width=\linewidth]{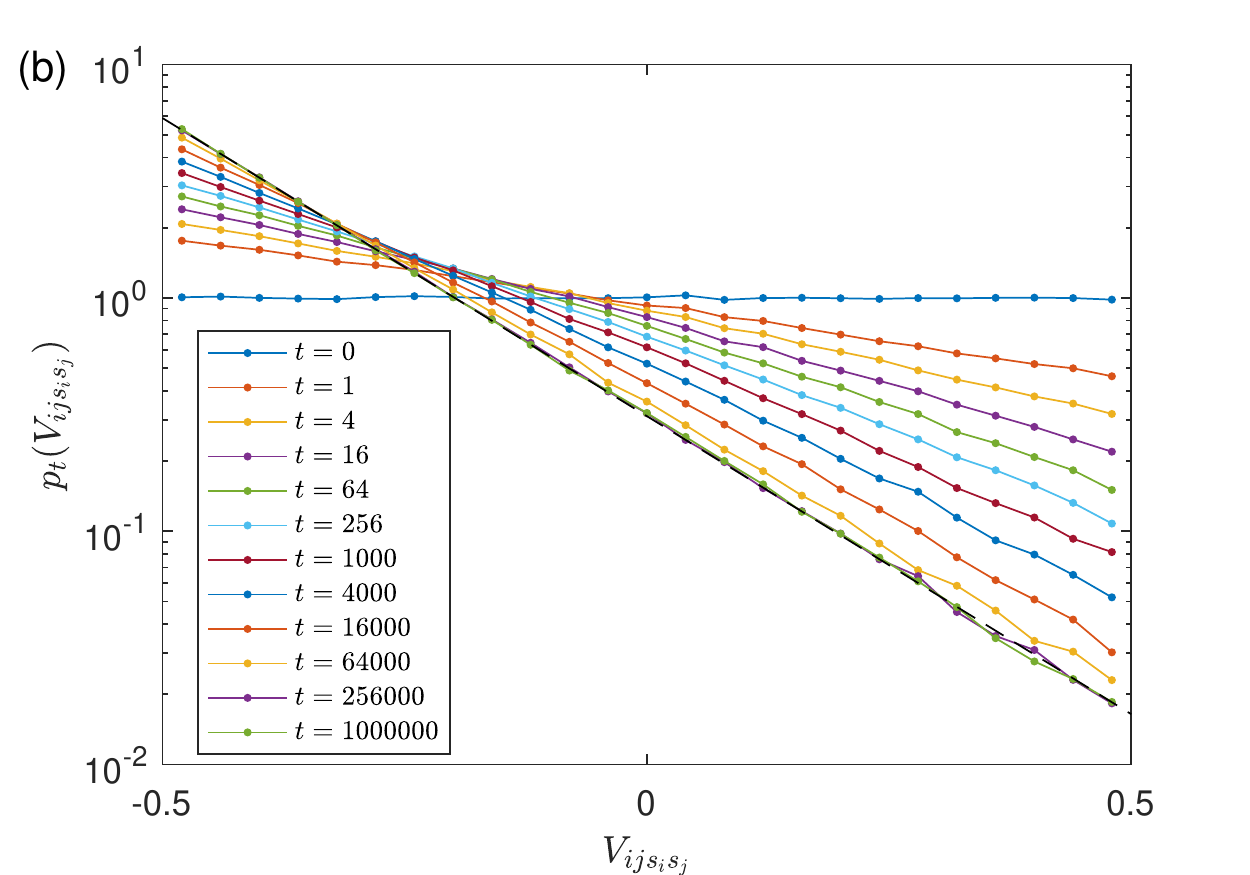}
\caption{(a) Plot of energy per particle $E/N$ against time $t$ from four independent runs adopting elementary (solid curves) and direct (dashed curves) initialization algorithms. The black dotted line shows $\av{E}/N$ from \eq{Eav}. (b) A semi-log plot of the probability distribution $p_t(\Vij)$ of realized interaction $\Vij$ at time $t$ from simulations adopting the  elementary initialization algorithm. $p_t(\Vij)$ for $t\ge 256000$ has converged to $p_{eq}$ from \eq{peq} indicated by the black dashed line.
For both (a) and (b), $T=0.170$ and $\phi_v=0.01$. 
}
\label{fig:E}
\end{figure}

\subsection{Elementary initial thermalization}
\label{sec:init1}
A straightforward approach is to generate each $\Vkl$ independently from the probability distribution $g$ taking into account the symmetry $V_{ijkl}=V_{jilk}$.  The system state is initialized at infinite temperature by putting each particle randomly onto an unoccupied site in the $L\times L$ lattice with uniform probability. This gives $n_i$ and $s_i$ at time $t=0$.
Thermalization kinetic Monte Carlo steps, typically performed using the rejection-free method explained above, are then conducted at the target temperature $T$ until equilibrium is attained.

During thermalization, equilibrium is indicated by the stabilization of various statistical measures such as the average particle energy $E/N$. 
We have checked numerically that equilibration can be performed successfully under various conditions and a particularly demanding example in the glassy phase is illustrated in Fig. \ref{fig:E}(a). 
The solid curves show $E/N$ against $t$  from two typical runs for $T=0.17$ and $\phiv=0.01$. They stabilize towards the average equilibrium value $\av{E}/N$ given in \eq{Eav}. 
Figure \ref{fig:E}(b) shows the evolution of the probability distribution $p_t(\Vij)$ of the realized interaction $\Vij$ at time $t$ from the same runs. It crossovers smoothly from the initial distribution $g$ toward the equilibrium distribution $p_{eq}$ in \eq{peq}. From both Figs. \ref{fig:E}(a) and (b), the system can be deemed equilibrium for the given $T$ and $\phiv$ for $t\agt 2.5\times 10^5$.
The results verify numerically \eq{Eav} and \eq{peq}. More importantly, they hence also verify the agreement between quenched and annealed averages used in their derivations. 

\subsection{Direct initialization method}
\label{sec:init}
System equilibration at large $N$ using thermalization Monte Carlo steps explained above can take very long runtime at low $T$. This is a major difficulty for many lattice models and most MD simulations of glass. Being able to directly construct equilibrium states is thus a highly desirable property.
This is possible for KCM with trivial energetics, non-spatial models \cite{derrida1980,ritort1995} and some frustrated spin models defined on triangular or related lattices \cite{newman1999}. It is also possible for MD simulations of a system with long-range shifted interactions \cite{mari2011,charbonneau2014}.
DPLM is in our knowledge the only finite-dimensional and energetically non-trivial lattice model of glass defined on a general lattice with this capability. 

First, we calculate the particle occupancy $n_i$ and the particle index $s_i$ at every site $i$. We start by simulating a simple identical-particle lattice gas because of the equivalent particle statistics (see \eq{ZLG}). It is performed with a constant NN particle interaction energy $U$ given in \eq{U} and we use the same computer code for DPLM with $\Vkl$ reduced to the constant $U$. Similar to the elementary initial thermalization approach described above, we initialize the particle positions randomly and then equilibrate the simple lattice gas by a thermalization run. It is computationally very efficient because of the absence of glassification at arbitrary $T$. The thermalized particle positions give $s_i$ and $n_i$ at time $t=0$. Note that due to the identical-particle nature of this part of the simulation, only $n_i$ is of interest. The precise particle permutation as specified by $s_i$ for the occupied sites is irrelevant because all permutations are equally probable (see \eq{Peqs}).

Second, we randomly generate $\Vkl$ from the annealed ensemble, which is statistically identical to the quenched ensemble (see \eq{Zequal}). Specifically, We sample all unrealized $\Vkl$ from the distribution $g$ while realized interactions $\Vij$ appearing in the state $\s$ are sampled from the Boltzmann distribution $p_{eq}$ given by \eq{peq}. This completes the generation of an equilibrium state at $T$.

In Fig. \ref{fig:E}(a),  the two dashed curves show the particle energy $E/N$ from two typical runs using this direct initialization method. They support that the systems have attained equilibrium energy once constructed and this numerically verifies the method. 
Note that at finite $N$, interactions from this approach differ in principle from that based on the elementary method in \app{sec:init1} in which all interactions are sampled from $g$. 
Nevertheless, since the total number of $\Vkl$ is of order $N^3$ while the number of bonds is of order $N$, the fraction of realized interactions sampled from $p_{eq}$ is only of order $1/N^2$. The fraction thus approaches zero at large $N$ and this demonstrations the equivalence of the elementary and the direct methods. Even after using our two-step tabulation approximation in \eq{Vpermute}, the fraction increases to order $1/N$ and still vanishes for large $N$. 
This approach of determining the particle arrangement $n_i$ and $s_i$ before generating the interaction $\Vkl$ is closely analogous to a planting method in Refs. \onlinecite{mari2011,charbonneau2014}.

\subsection{Software reliability}
Correct software implementation is highly nontrivial because minor programming mistakes may affect the particle dynamics only occasionally and can be very difficult to spot. One helpful consistency check is to measure the probability distribution of the interaction energy $\Vij$ at equilibrium and compare with the exact distribution in \eq{peq}. We have also conducted more general Boltzmann distribution tests \cite{lam2008} by  performing long simulations using a small lattice with all but several particles frozen. Then, only a few thousand different configurations will be realized. We measure the total occurrence durations and the system energies of all these configurations and make sure that the results agree with the Boltzmann distribution within the expected statistical errors. With these tests, we believe that our software implementation is highly reliable.

\bibliography{polymer_short}
\end{document}